\documentclass[prd,aps,showpacs,nofootinbib,preprint,eqsecnum]{revtex4}
\usepackage{amsmath} 
\usepackage{amssymb}
\usepackage{amsfonts}
\usepackage{graphicx,bm}
\usepackage{dcolumn}
\usepackage{color,amsxtra}
\usepackage{epsf}
\usepackage{enumerate}
\usepackage{hhline}
\usepackage{array}
\usepackage{tabularx}
\usepackage{subfigure}

\usepackage{fancyhdr}
\usepackage{mathrsfs}

\newcommand{\be}{\begin{equation}}
\newcommand{\ee}{\end{equation}}
\newcommand{\bea}{\begin{eqnarray}}
\newcommand{\eea}{\end{eqnarray}}
\newcommand{\beaa}{\begin{eqnarray*}}
\newcommand{\eeaa}{\end{eqnarray*}}

\newcommand{\e}{\mathrm{e}}

\allowdisplaybreaks[4]

\newcommand{\Eqn}[1]{&\hspace{-0.2em}#1\hspace{-0.2em}&}
\def\Vec#1{\mbox{\boldmath $#1$}}

\def\be{\begin{equation}}
\def\ee{\end{equation}}
\def\bea{\begin{eqnarray}}
\def\eea{\end{eqnarray}}

\def\e{\mathrm{e}}

\begin{document}

\title{Inflationary cosmology in modified gravity theories} 

\author{Kazuharu Bamba$^{1, 2}$ and Sergei D. Odintsov$^{3, 4}$ 
}
\affiliation{
$^1$Leading Graduate School Promotion Center,
Ochanomizu University, Tokyo 112-8610, Japan\\
$^2$Department of Physics, Graduate School of Humanities and Sciences, Ochanomizu University, Tokyo 112-8610, Japan\\ 
$^3$Consejo Superior de Investigaciones Cient\'{\i}ficas, ICE/CSIC-IEEC, 
Campus UAB, Facultat de Ci\`{e}ncies, Torre C5-Parell-2a pl, E-08193
Bellaterra (Barcelona), Spain\\
$^4$Instituci\'{o} Catalana de Recerca i Estudis Avan\c{c}ats
(ICREA), Passeig Llu\'{i}s Companys, 23 08010 Barcelona, Spain
}


\begin{abstract} 
We review inflationary cosmology in modified gravity such as $R^2$ gravity with its extensions in order to generalize the Starobinsky inflation model. 
In particular, we explore inflation realized by three kinds of effects: 
modification of gravity, the quantum anomaly, and the $R^2$ term in loop quantum cosmology. It is explicitly demonstrated that in these inflationary models, the spectral index of scalar modes of the density perturbations and the tensor-to-scalar ratio can be consistent with the Planck results. 
Bounce cosmology in $F(R)$ gravity is also explained. 
\end{abstract}

\hspace{13.0cm} OCHA-PP-331

\maketitle

\def\thesection{\Roman{section}}
\def\theequation{\Roman{section}.\arabic{equation}}

\section{Introduction}
 
Inflation in the early universe has recently been studied much more 
extensively because of the BICEP2 experiment~\cite{Ade:2014xna} 
in terms of the primordial gravitational waves, 
in addition to the Wilkinson Microwave anisotropy probe (WMAP)~\cite{Spergel:2003cb,Spergel:2006hy,Komatsu:2008hk,Komatsu:2010fb,Hinshaw:2012aka} and the Planck satellite~\cite{Ade:2013lta,Ade:2013uln} on the unisotropy of the cosmic microwave background (CMB) radiation. For a standard inflationary scenario like chaotic inflation~\cite{Linde:1983gd}, the existence of the inflaton field is assumed, whose potential contributes to inflation. 

On the other hand, the accelerated expansion of the universe including inflation and the late-time acceleration, i.e., dark energy problem, can be realized in modified gravity theories such as $F(R)$ gravity (for reviews on inflation, 
see, e.g.,~\cite{Linde:2014nna,Gorbunov:2011zzc}, whereas for dark energy and modified gravity, 
see, for example, Refs.~\cite{Nojiri:2010wj,Nojiri:2006ri,Bamba:2013iga,Bamba:2014eea,Book-Capozziello-Faraoni,Capozziello:2011et,delaCruzDombriz:2012xy,Bamba:2012cp,Joyce:2014kja}). For instance, the trace-anomaly driven inflation such as the Starobinsky inflation~\cite{Staro,Alex} is well known. 

Indeed, the WMAP and Planck data~\cite{Spergel:2003cb,Spergel:2006hy,Komatsu:2008hk,Komatsu:2010fb,Hinshaw:2012aka,Ade:2013lta,Ade:2013uln} support 
a kind of the trace-anomaly driven inflation with the $R^2$ term. 
Such a theory can be regarded as modified gravity because the $R^2$ term or its higher derivative term of the trace-anomaly term $\Box R$, which leads to the long enough inflation and graceful exit from it~\cite{Staro}, is the effective action of gravity, where $\Box$ is the covariant d'Alembertian for 
a scalar quantity\footnote{In Refs.~\cite{Nojiri:2010pw,Bamba:2012ky,Zhang:2011uv}, the features of inflation in non-local gravity including such a non-local term as $\Box R$ has been analyzed in detail.}. 

In this paper, we review the main results in Refs.~\cite{Sebastiani:2013eqa,Bamba:2014jia,Amoros:2014tha}. 
The main purpose of this paper is to explain the recent developments 
on inflationary models to realize the Planck results 
in the so-called $R^2$ gravity (namely, the action consist of the Einstein-Hilbert term plus $R^2$ term) with further extensions, which can be regarded as 
a kind of $F(R)$ gravity. Particularly, we consider inflation (i) derived by 
modification terms of gravity~\cite{Sebastiani:2013eqa}, (ii) through the quantum anomaly~\cite{Bamba:2014jia}, 
and (iii) in $R^2$ gravity in the framework of the so-called 
loop quantum cosmology (LQC)~\cite{Bojowald:2001xe,bo02a,Ashtekar:2005qt,Ashtekar:2007tv,Ashtekar:2007em} to 
include quantum effects~\cite{Amoros:2014tha} 
(for reviews on LQC, see, for example,~\cite{Thiemann:2002nj,abl03,Bojowald:2006da,t01,Bojowald:2008zzb,as11,Bojowald:2012xy}). 
In addition, we state the recent progress of the bounce cosmology 
in $F(R)$ gravity by presenting the important consequences in Refs.~\cite{Odintsov:2014gea,Odintsov:2015uca}. 
We use units of $k_\mathrm{B} = c = \hbar = 1$ and express the
gravitational constant $8 \pi G_{\mathrm{N}}$ by
${\kappa}^2 \equiv 8\pi/{M_{\mathrm{Pl}}}^2$ 
with the Planck mass of $M_{\mathrm{Pl}} = G_{\mathrm{N}}^{-1/2} = 1.2 \times 
10^{19}$\,\,GeV. 

The organization of the paper is the following. 
In Sec.~II, we study inflation induced by modification of gravity. 
In Sec.~III, we explore the trace-anomaly driven inflation 
in modified gravity. 
In Sec.~IV, we investigate $R^2$ gravity in the context of LQC. 
Furthermore, in Sec.~V, we reconstruct $F(R)$ gravity to 
realize the cosmological bounce in LQC.
Finally, conclusions are described in Sec.~VI. 

\section{Inflation induced by modification of gravity} 

In this section, we review inflation in modified gravity, 
particularly $F(R)$ gravity, based on Ref.~\cite{Sebastiani:2013eqa}. 
The deviation of $F(R)$ gravity from 
general relativity may be interpreted as a kind of quantum corrections in the early universe, or such a modification of gravity could be motivated by the so-called ultraviolet (UV) completion of quantum gravity. 
In fact, the Starobinsky inflation~\cite{Staro} can be regarded as 
inflation induced by the modification term of $R^2$ from general relativity. 
We here attempt to examine inflation by the other forms of modification of 
gravity.

\subsection{Conformal transformation}  

We first explain the conformal transformation from 
$F(R)$ gravity in the Jordan frame 
to the corresponding scalar field theory 
in the Einstein frame~\cite{Bamba:2012cp,Bamba:2008hq}. 
The action of $F(R)$ gravity is represented as 
$S = \int d^4 x \sqrt{-g} \left[F(R)/\left(2\kappa^2\right) \right]$, 
where $g$ is the determinant of the metric tensor $g_{\mu\nu}$. 
We use a conformal transformation 
$\hat{g}_{\mu\nu} = \Omega^{-2} g_{\mu\nu}$ with 
$\Omega^2 \equiv F_{R}$, 
where the hat denotes quantities in the Einstein frame, and 
the subscription of $F_R$ denotes the derivative with respect to 
$R$ as $F_R (R) \equiv d F(R)/d R$. 
Here, we introduce a scalar field 
$\varphi \equiv -\sqrt{3/2} \left(1/\kappa\right) \ln F_R$. 
Through the conformal transformation,  
the action in the Einstein frame reads~\cite{Maeda:1988ab,F-M} 
\begin{eqnarray} 
S_{\mathrm{E}} \Eqn{=} 
\int d^4 x \sqrt{-\hat{g}} \left( \frac{\hat{R}}{2\kappa^2} - 
\frac{1}{2} \hat{g}^{\mu\nu} {\partial}_{\mu} \varphi {\partial}_{\nu} \varphi 
- V(\varphi) \right)\,,
\label{eq:2.1}\\ 
V(\varphi) \Eqn{=} \frac{F_R \hat{R}-F}{2\kappa^2 \left(F_R \right)^2}\,, 
\label{eq:2.2}
\end{eqnarray}
with $F_R = \exp \left(-\sqrt{2/3} \kappa \varphi\right)$. 
This is the action for a canonical scalar field $\varphi$ with its potential 
$V(\varphi)$. 
For the Starobinsky inflation model~\cite{Staro} with 
$F(R) = R + \alpha_\mathrm{S} \kappa^2 R^2$, 
where $\alpha_\mathrm{S}$ is a constant, we have $V(\varphi) = \left[1/\left(8\alpha_\mathrm{S} \kappa^2\right)\right] \left( 1-\exp\left(-\sqrt{2/3} \kappa \varphi \right) \right)^2$.

\subsection{Slow-roll inflation} 

We describe the procedures to deal with the so-called slow-roll inflation. 
We consider the action in Eq.~(\ref{eq:2.1}) and regard $\varphi$ as the inflaton field. 
We suppose the flat Friedmann-Lema\^{i}tre-Robertson-Walker (FLRW) metric 
$ds^2 = - dt^2 + a^2(t) d\Vec{x}^2$. 
Here, $a(t)$ is the scale factor. 
The Hubble parameter is defined as 
$H \equiv \dot{a}/a$, where the dot denotes the time derivative. 
The gravitational field equations in this background read 
$3H^{2}/\kappa^{2} = \dot{\varphi}^2/2 + V(\varphi)$ and 
$-\left( 2\dot H+3H^{2} \right)/\kappa^{2} = \dot{\varphi}^2/2 - V(\varphi)$. 
Moreover, the equation of motion (EoM) for $\varphi$ becomes 
$\ddot{\varphi} + 3H\dot{\varphi} + dV(\varphi)/d\varphi = 0$. 

For the slow-roll regime, we impose the slow-roll approximations of 
$\dot{\varphi}^2/2 \ll V(\varphi)$ on the Friedmann equation and $\left|\ddot{\varphi}\right| \ll \left|3H\dot{\varphi}\right|$ on the EoM for $\varphi$, so that we can find $3H^{2}/\kappa^{2} \approx V(\varphi) 
\approx \mathrm{constant}$ and $3H\dot{\varphi} + dV(\varphi)/d\varphi \approx 0$. Furthermore, we define the slow-roll parameters 
$\epsilon \equiv -\dot{H}/H^2 = \left[1/\left(2\kappa^2\right)\right] 
\left[\left(dV(\varphi)/d\varphi \right)/V(\varphi)\right]^2 \, (\ll 1)$ 
and $\eta \equiv -\ddot{H}/\left( 2H\dot{H}\right) = \left[1/\kappa^2\right] \left[\left(d^2V(\varphi)/d \varphi^2 \right)/V(\varphi)\right] \, (\ll 1)$. 
During inflation, these parameters should be much smaller than unity. 
In addition, the number of $e$-folds is 
$N_e \equiv \ln \left(a_\mathrm{f}/a_\mathrm{i}\right) 
= \int_{t_\mathrm{i}}^{t_\mathrm{f}} H dt \approx 
\kappa^2 \int_{\varphi_\mathrm{f}}^{\varphi_\mathrm{i}} 
\left[V/\left(dV(\varphi)/d\varphi\right)\right] d\varphi$. 
Here, $a_\mathrm{i}$ ($\varphi_\mathrm{i}$) and $a_\mathrm{f}$ ($\varphi_\mathrm{f}$) are the values of the scale factor $a$ (the scalar field $\varphi$) at 
the initial time $t_\mathrm{i}$ and end of time $t_\mathrm{f}$ of inflation, respectively. Moreover, in deriving the second approximate equality, we have used 
the second gravitational field equation with the slow-roll approximation. 
The amplitude of the power spectrum for the curvature perturbations 
is expressed as  
$\Delta_{\mathcal R}^2=\kappa^2 H^2/\left(8\pi^2\epsilon\right) \approx 
\kappa^4 V/\left(24\pi^2\epsilon\right)$, where 
in the second approximate equality follows from the 
Friedmann equation operated the slow-roll approximation, 
and $n_\mathrm{s}$ and $r$ are written as 
$n_\mathrm{s} - 1= -6\epsilon+2\eta$ and 
$r=16\epsilon$~\cite{Mukhanov:1981xt, L-L}. 
The detailed explanations on the reconstruction of potential of inflationary models has been executed in Ref.~\cite{Lidsey:1995np}.

\subsection{Reconstruction of $F(R)$ gravity}  

There are two possible ways to reconstruct $F(R)$ gravity models to 
realize inflation. 
One is to start from the action in the Einstein frame. 
The other is to reconstruct the action in the Jordan frame. 
In this subsection, we consider the former way. 
We also explain the latter way in the next subsection. 

The main purpose of our investigations is that we study the generalization of the Starobinsky inflation model. 
To execute it, we take an appropriate form of $V(\varphi)$ in the Jordan frame, which is an extended form from that in the Starobinsky inflation model. 
By taking the derivative of Eq.~(\ref{eq:2.2}) with respect to $R$, 
we find 
\begin{equation} 
R F_R =-\sqrt{6}\kappa 
\frac{d}{d\varphi} \left(V(\varphi)\exp \left(-2\sqrt{\frac{2}{3}}\kappa\varphi\right) \right)\,. 
\label{eq:2.3}
\end{equation}
Through this equation, i.e., Eq.~(\ref{eq:2.2}), we reconstruct the form 
of $F(R)$ in the Jordan frame from the potential $V(\varphi)$ in the Einstein frame.

\subsubsection{Extension of the Starobinsky inflation model}

As the simplest model, we explore the following potential
\begin{equation} 
V(\varphi)=c_1+c_2\exp\left( \sqrt{\frac{2}{3}} \kappa \varphi\right) +c_3
\exp\left( 2\sqrt{\frac{2}{3}} \kappa \varphi\right) \,, 
\label{eq:2.4}
\end{equation}
where $c_1 (\neq 0)$, $c_2$, and $c_3$ are constants. In this case, from Eq.~(\ref{eq:2.3}) we have $2c_1 F_R^2 +c_2 F_R - RF_R=0$. By solving this equation with $F_R \neq 0$, 
we eventually find that the corresponding form of $F(R)$ can be 
expressed as 
\begin{equation} 
F(R)=R+\frac{R^2}{4c_1}+ c_1 -c_3\,. 
\label{eq:2.5}
\end{equation}
Here, we have set $-c_2/\left(2c_1\right) = 1$ in order to reproduce the Einstein-Hilbert term in $F(R)$ and 
used Eq.~(\ref{eq:2.2}) in determining the integration constant. 
If $c_1 = c_3$ (which leads to $c_3 = -c_2/2$), 
this model is equivalent to the Starobinsky inflation model 
with $V(\varphi) = c_1 \left( 1-\exp\left(-\sqrt{2/3} \kappa \varphi \right) \right)^2$, whereas for $c_1 > c_3$, this corresponds to an extended model of the 
Starobinsky inflation with a cosmological constant. 
As the possibile origins of 
Such a cosmological constant emerging at the large curvature regime 
could originate from the quantum effects, or a modification term of gravity 
removing the cosmological constant at the small curvature regime~\cite{Cognola:2008,Linder:2009,Elizalde:2011,Oikonomou:2013rba}. 

In the following, we set $c_3 = 0$ for simplicity and introduce a positive $\gamma_1 \, (>0)$ to express $c_1$ as $c_1 = \gamma_1/\left(4\kappa^2 \right)$. 
Here, $\gamma_1$ has the mass dimension $2$ and the dimensionless quantity 
$\gamma_1/M_\mathrm{Pl} \ll 1$, so that in the higher-curvature regime, the correction to the Einstein gravity can appear. 
We explore the inflationary dynamics in this extended model with the potential 
$V(\varphi) = \gamma_1/\left(4\kappa^2 \right) - \left[\gamma_1/\left(2\kappa^2 \right) \right] \exp\left( \sqrt{2/3} \kappa \varphi\right)$. 
We consider the case that the inflaton slowly rolls from the initial value 
with its large negative amplitude down to the minimum of the potential 
as $V(\varphi = 0) = -\gamma_1/\left(4\kappa^2 \right) \, (<0)$. 
In this case, from the gravitational field equations we find that 
the exponential inflation can be realized as $a(t) = a_\mathrm{i} \exp \left( H_\mathrm{inf} t \right)$ with the Hubble parameter $H_\mathrm{inf} \equiv 
\left(1/2\right) \sqrt{\gamma_1/3}$ during inflation and 
$a_\mathrm{i}$ a constant. Furthermore, the solution of $\varphi$ reads $\varphi = -\sqrt{3/2} \kappa \ln\left[ \left(1/3\right) \sqrt{2\gamma_1/3} \left(t_\mathrm{i} -t \right) \right]$. Around the beginning of inflation $t \simeq t_\mathrm{i}$, $|\varphi| \gg 1$ and the slow-roll parameters are $\epsilon =\left(4/3 \right)\left[1-\exp\left(-\sqrt{2/3} \kappa \varphi \right) \right]^{-2} \ll 1$ and $|\eta| = \left(4/3 \right)\left|1-\exp\left(-\sqrt{2/3} \kappa \varphi \right) \right|^{-1} \ll 1$. 
These slow-roll parameters become of order of unity when $\varphi$ approaches 
$\varphi_\mathrm{f} \approx -0.17\sqrt{3/2}/\kappa$. 
For $\left|\varphi_\mathrm{i}\right| \gg \left|\varphi_\mathrm{f}\right|$, 
the number of $e$-folds is given by 
$N_e \approx \left(1/2\right)\sqrt{\gamma_1/6}t_\mathrm{i}$. 
In addition, we find $t_\mathrm{f} = t_\mathrm{i} - 3\sqrt{3/\left(2\gamma_1\right)}\exp\left(0.17\right)$. 
For $N_e = 60$, we have $\varphi_\mathrm{i} \approx 1.07 M_\mathrm{Pl}$. 
The slow-roll parameters are also represented as 
$\epsilon \approx 3/\left(4N_e^2\right)$ and $\left|\eta\right| \approx 1/N$. 
As a consequence, we obtain
\begin{equation}
\Delta_{\mathcal R}^2 \approx \frac{\kappa^2\gamma_1 
N_e^2}{72\pi^2} \,,
\quad 
n_\mathrm{s} - 1 \approx -\frac{2}{N_e}  \,,
\quad
r \approx \frac{12}{N_e^2} \,. 
\label{eq:2.6} 
\end{equation}
Here, we remark that $\Delta_{\mathcal R}^2 \approx \kappa^2\gamma_1 
N_e^2/\left(72\pi^2\right) \ll \kappa^2 M_\mathrm{Pl} N_e^2/\left(72\pi^2\right)$ because $\gamma_1 \ll M_\mathrm{Pl}$. 

The observations obtained from the Planck satellite suggest 
$n_{\mathrm{s}} = 0.9603 \pm 0.0073\, (68\%\,\mathrm{CL})$ and 
$r < 0.11\, (95\%\,\mathrm{CL})$~\cite{Ade:2013lta}. 
In this model, for $n_\mathrm{s} < 1$ and $r<0.11$, 
we see that $n_\mathrm{s} >1-\sqrt{0.11/3} = 0.809$. 
Accordingly, for $N_e = 60$, we acquire $n_\mathrm{s}=0.967$ and 
$r=3.00 \times 10^{-3}$. 
Thus, in this model, the spectral index $n_\mathrm{s}$ of the curvature 
perturbations and the tensor-to-scalar ratio $r$ consistent with 
the Planck result can be realized. 
Various descriptions of 
inflationary models in terms of scalar field models~\cite{Bamba:2014daa} and 
perfect fluid as well as $F(R)$ gravity~\cite{Bamba:2014wda} 
have been examined. 
Moreover, the effects of quantum corrections on inflation have been explored in Refs.~\cite{Bamba:2014mua,Cognola:2014pha,Rinaldi:2014gha}. 
We note that the BICEP2 experiment has recently detected the $B$-mode polarization of the cosmic microwave background (CMB) radiation with the tensor to scalar ratio $r = 0.20_{-0.05}^{+0.07}\, (68\%\,\mathrm{CL})$~\cite{Ade:2014xna}. 
There have been proposed several discussions on the method to obtain this result regarding the subtraction of the foreground data, e.g., Refs.~\cite{Ade:2014gna,Ade:2014zja,Adam:2014oea,Mortonson:2014bja,Kamionkowski:2014}. A study to support the BICEP2 results has also been reported in Ref.~\cite{Colley:2014nna}. 
Very recently, 
the collaboration between BICEP2/Keck and Planck 
has released the result of  $r < 0.12$ for the wave number $k = 0.05 \, \mathrm{Mpc}^{-1}$ of tensor mode of the density perturbations~\cite{BICEP2-Keck-Planck}.

\subsubsection{Power-law corrections to general relativity}

Next, we examine the following potential. 
\begin{eqnarray}
V(\varphi) \Eqn{=} \frac{1}{2\kappa^2}\left\{\left(\frac{1}{\beta 
q}\right)^{1/\left(q-1\right)}\left(1-\exp\left( \sqrt{\frac{2}{3}}\kappa\varphi \right) 
\right)^{q/\left(q-1\right)}
\exp\left[ \left( \frac{q-2}{q-1} \right) \sqrt{\frac{2}{3}}\kappa\varphi 
\right] 
\left(\frac{q-1}{q}\right)
\right. 
\nonumber\\
&&\left.+R_\mathrm{c} 
\exp \left( \sqrt{\frac{2}{3}} \kappa\varphi \right)
\left(\exp\left( \sqrt{\frac{2}{3}}\kappa\varphi \right) 
-1\right)
-\Lambda_\mathrm{p}
\exp\left( 2\sqrt{\frac{2}{3}}\kappa\varphi \right) 
\right\}\,.
\label{eq:2.7} 
\end{eqnarray} 
For this potential in the Einstein frame, 
a model in which a generic power-law correction term is added to the Einstein-Hilbert term is reconstructed as 
\begin{equation}
F(R)=R+\beta \left(R+R_\mathrm{c}\right)^q+\Lambda_\mathrm{p} \,. 
\label{eq:2.8}
\end{equation}
Here, $\beta \, (>0)$ is a dimensionful positive constant, 
$R_\mathrm{c}$ and $\Lambda_\mathrm{p}$ are constant, 
and $q>1$ ($q \neq 2$). 
Inflationary models in such a power-law type gravity with 
$q \lesssim 2$ 
has also been examined in Ref.~\cite{Motohashi:2014tra}. 
Through the same procedures as those executed for the previous case in Eqs.~(\ref{eq:2.4}) and (\ref{eq:2.5}), namely, by deriving 
the Hubble parameter at the inflationary stage, the scale factor, 
the solution of the inflaton $\varphi$, the slow-roll parameters, 
and the number of $e$-folds, if $n$ is close to $2$, we find
\begin{eqnarray}
&&
\Delta_{\mathcal R}^2 \approx \frac{\left(q-1\right)^3
}{16\pi^2 q\left(2-q\right)^2}
\kappa^2 
\exp\left[ \frac{2}{3}N_e\frac{\left(q-2\right)^2}{\left(q-1\right)^2} \right]
\left(\frac{1}{\beta q}\right)^{1/\left(q-1\right)}\,, 
\nonumber \\
&&
n_\mathrm{s} - 1 \approx -\frac{8\left(2-q\right)}{3\left(q-1\right)}\,,\quad 
r \approx \frac{16\left(2-q\right)^2}{3\left(q-1\right)^2}\,. 
\label{eq:2.9}
\end{eqnarray}
Indeed, for $q =1.99$, we find $n_\mathrm{s}=0.962$ and 
$r=1.08 \times 10^{-3}$. Therefore, this inflationary model can be compatible with the Planck analysis.

\subsection{Reconstruction method of $F(R)$ gravity in the Jordan frame}  

In this subsection, we reconstruct the form of $F(R)$ in the Jordan frame. 
We note that cosmology in the Einstein frame may differ from that in the Jordan frame due to their physical non-equivalence. 
Hence, it is more convenient to consider these theories in the Einstein 
and Jordan frames as different cosmological theories. Here, we discuss inflation in $F(R)$ gravity without its transformation to scalar-tensor theory. 
Eventually, the results may be different. Nevertheless, we demonstrate that $F(R)$ inflation is also consistent with the observations by the Planck satellite. 
{}From the other point of view, for the fairness, 
it should also be remarked that there are the debates on the issue of the equivalence between the (Jordan and Einstein) conformal frames in Refs.~\cite{Prokopec:2012ug,George:2013iia,Kaiser:1995nv}. 
Especially, the investigations in Ref.~\cite{Kaiser:1995nv} seems to support equivalence 
of conformal frames for inflationary scenarios. It 
is an issue of presentation rather than substance, but in the 
interest of fairness other points of view should be 
mentioned. 

The reconstruction method of $F(R)$ gravity proposed in Ref.~\cite{Nojiri:2009kx} is as follows (for another reconstruction method of $F(R)$ gravity, see Refs.~\cite{Nojiri:2006gh,Nojiri:2006be,Nojiri:2011kd,delaCruzDombriz:2006fj}). We consider the action of $F(R)$ gravity with matter action $S_\mathrm{matter}$ as $S = \int d^4 x \sqrt{-g} \left[F(R)/\left(2\kappa^2\right) \right] + S_\mathrm{matter}$. We define the number of $e$-folds as $\bar{N} \equiv \ln \left(a_*/a\right)$ with $a_*$ the scale factor 
at the fiducial time $t_*$. We define $\bar{G}(\bar{N}) \equiv H^2(\bar{N})$, 
so that $R$ can be expressed as $R(\bar{N}) = 3\left[\left(d\bar{G}(\bar{N})/dN\right)+4\bar{G}(\bar{N})\right]$. 
By solving this equation inversely, we get $\bar{N} = \bar{N}(R)$. 
In the flat FLRW space-time, the Friedmann equation 
can represented as the second order differential equation of $F(R)$ 
with respect to $R$, given by
\begin{eqnarray}
&& -9\bar{G}(\bar{N}(R))\left( 4\bar{G}_{\bar{N}} (\bar{N}(R))+G_{\bar{N} \bar{N}} (\bar{N}(R)) \right)F_{RR}(R)
\nonumber \\ 
&& 
{}+
3\left(\bar{G}(\bar{N}(R))+\frac{1}{2}G_{\bar{N}}(\bar{N}(R)) \right)F_R(R)-\frac{F(R)}{2}+\kappa^2\rho_\mathrm{matter}=0\,,
\label{eq:2.10} 
\end{eqnarray}
where $F_{RR} \equiv d^2F(R)/dR^2$, $\bar{G}_{\bar{N}} \equiv d\bar{G}(\bar{N}(R))/d\bar{N}$, $\bar{G}_{\bar{N} \bar{N}} \equiv d^2\bar{G}(\bar{N}(R))/d\bar{N}^2$, and $\rho_\mathrm{matter}$ is the energy density of matter. 

As an example found in Ref.~\cite{Bamba:2014wda}, 
we study an exponential form 
$\bar{G}_{\bar{N}} = H^2 (\bar{N})= \bar{G}_1 \e^{\tau \bar{N}} 
+ \bar{G}_2$, where $G_1 \,(<0)$, $G_2 \,(>0)$, and $\tau \,(>0)$ 
are constants. For this expression, we have 
$\e^{\tau N}=\left(R-12\bar{G}_2\right)/\left[3\bar{G}_1 \left(4+\tau\right)\right]$. When the matter contribution is negligible, namely, 
$\rho_\mathrm{matter} = 0$, the solution of Eq.~(\ref{eq:2.10}) 
is derived as 
\begin{equation}
F(R)=Q_1 F\left(\omega_{+},\omega_{-},l;\vartheta\right)+Q_2\left(12\bar{G}_2-R\right)^{\left(1+1/\tau \right)} F\left(1+\omega_{-}+\frac{1}{\tau},1+\omega_{+}+\frac{1}{\tau},2+\frac{1}{\tau};\vartheta\right) \,. 
\label{eq:2.11}
\end{equation}
Here, $\omega_{\pm}$ and $\vartheta$ are defined as 
\bea
\omega_{\pm} \equiv \frac{-3 \tau -2\pm\sqrt{\tau^2-20 \tau +4}}{4 \tau}\,, 
\quad \vartheta \equiv \frac{12\bar{G}_2-R}{3\bar{G}_2\left(4+\tau\right)}\,, 
\label{FRNe12-2}
\eea
with $F(\varsigma_1, \varsigma_2, \varsigma_3; \vartheta)$ 
the hypergeometric function, where $\varsigma_i$ ($i=1, \dots, 3$) 
are constants. 
If $(\bar{N}, G_1, G_2) = (50.0, -1.10, 10.0)$ and 
$(60.0, -1.20, 15.0)$, we obtain 
$(n_\mathrm{s}, r) = (0.963, 6.89 \times 10^{-2})$ 
and $(0.965, 5.84 \times 10^{-2})$, respectively\footnote{
The running of the spectral index 
$\alpha_\mathrm{s} \equiv d n_\mathrm{s}/d \ln k$ 
is also estimated as $\alpha_\mathrm{s} = -5.06 \times 10^{-5}$ 
and $-4.51 \times 10^{-5}$ for $(\bar{N}, G_1, G_2) = (50.0, -1.10, 10.0)$ and 
$(60.0, -1.20, 15.0)$, respectively.}. 
Therefore, this model can yield the values of $n_\mathrm{s}$ and $r$ 
indicated by the Planck analysis.

\section{Trace-anomaly driven inflation in modified gravity} 

In this section, we review inflation by the quantum anomaly in the framework of $F(R)$ gravity by following Ref.~\cite{Bamba:2014jia}. 
The effect of the trace anomaly on inflation in 
$F(T)$ gravity with $T$ the torsion scalar in teleparallelism 
has also been studied in Ref.~\cite{Bamba:2014zra} (the explanations of teleparallelism exist, e.g., in Refs.~\cite{Bamba:2014eea,Bamba:2012cp}).

\subsection{Quantum anomaly}

It is known that the quantum anomaly appears via 
the procedure of the renormalization. 
For four-dimensional space-time, 
the trace of the energy momentum tensor $T_{\mu\nu}^{(\mathrm{QA})}$ originating from the quantum anomaly 
becomes~\cite{Deser:1976yx,Duff:1977ay,Birel,Duff:1993wm,Nojiri:1998dh}
\begin{eqnarray}
\left\langle \, T_{\mu}^{(\mathrm{QA}) \mu} \, \right\rangle \Eqn{=} \alpha_1 \left(\mathcal{W}+\frac{2}{3}\Box R\right)- \alpha_2 \mathcal{G} + \alpha_3 \Box R\,,
\label{eq:3.1} \\
\mathcal{W} \Eqn{\equiv} \mathcal{C}^{\mu\nu\rho\sigma}\mathcal{C}_{\mu\nu\rho\sigma} 
= R^{\mu\nu\rho\sigma}R_{\mu\nu\rho\sigma}-2R^{\mu\nu}R_{\mu\nu}+\frac{1}{3}R^2\,,\label{eq:3.2} \\
\mathcal{G} \Eqn{\equiv} R^{\mu\nu\rho\sigma}R_{\mu\nu\rho\sigma}-4R^{\mu\nu}R_{\mu\nu}+R^2\,. 
\label{eq:3.3} 
\end{eqnarray}
Here, the brackets $\langle \, \rangle$ denotes the vacuum expectation value. 
Moreover, $R_{\mu\nu\rho\sigma}$ is the Riemann tensor, $R_{\mu\nu}$ is the Ricci tensor, $R$ is the scalar curvature, $\mathcal{C}_{\mu\nu\rho\sigma}$ is the Weyl tensor, to whose square $\mathcal{W}$ corresponds, $\mathcal{G}$ is the Gauss-Bonnet invariant, and 
$\Box = g^{\mu\nu} \nabla_{\mu}\nabla_{\nu}$ with ${\nabla}_{\mu}$ the covariant derivative associated with the metric tensor $g_{\mu\nu}$ is the covariant d'Alembertian. 
In addition, the coefficients are defined as  
$\alpha_1 \equiv \left(N_\mathrm{S}+6N_\mathrm{F}+12N_\mathrm{V}\right)/\left(1920\pi^2\right)$, $\alpha_2 \equiv \left(N_\mathrm{S}+11N_\mathrm{F}+62N_\mathrm{V}\right)/\left(5760\pi^2\right)$, 
$\alpha_3 \equiv -N_\mathrm{V}/\left(96\pi^2\right)$ 
with the number of real scalar fields $N_\mathrm{S}$, 
that of the Dirac (fermion) fields $N_\mathrm{F}$, 
and that of vector fields $N_\mathrm{V}$, 
where we have neglected the contributions from gravitons and higher-derivative conformal scalars. 

We set the values of $\alpha_1$ and $\alpha_2$ positive, but the qualitative consequences do not depend on $\alpha_1$, $\alpha_2$, and $\alpha_3$. 
For example, in the $\mathcal{N}=4$ SU(N) super Yang-Mills theory, we have $\alpha_1=\alpha_2=\bar{N}^2/\left(64\pi^2\right) \, (> 0)$ and 
$\alpha_3=-\bar{N}^2/\left(96\pi^2\right)$, where we have used 
$N_\mathrm{S} = 6 \bar{N}^2$, $N_\mathrm{F} = 2\bar{N}^2$, and $N_\mathrm{V} = \bar{N}^2$ with $\bar{N} \gg 1$. Here, $\left(2/3\right)\alpha_1 + 
\alpha_2 = 0$. 
However, if the action has an additional $R^2$ term as~\cite{Haw} 
$\left[ \alpha_4 \bar{N}^2/\left(192\pi^2\right) \right] 
\int d^4 x \sqrt{-g}\,R^2$, where $\alpha_4 \, (>0)$ is a positive constant, 
we find $\left(2/3\right)\alpha_1 + 
\alpha_2 = -\alpha_4 \bar{N}^2/\left(16\pi^2\right)$. 
In the classical level, the vacuum expectation value of $\left\langle \, T_{\mu\nu}^{(\mathrm{QA})} \, \right\rangle$ in Eq.~(\ref{eq:3.1}) can be regarded as a contribution of matter in the right-hand side as $R_{\mu\nu} -\left(1/2\right)g_{\mu\nu} R = \kappa^2 \left\langle \, T_{\mu\nu}^{(\mathrm{QA})} \, \right\rangle$. 
Its trace reads $R = \alpha_1 \left[\mathcal{W}+ \left(2/3\right) \Box R \right] - \alpha_2 \mathcal{G} + \alpha_3 \Box R + \left[ \alpha_4 \bar{N}^2 \kappa^2/\left(16\pi^2\right) \right] \Box R$. 
Accordingly, for the Yang-Mills theory in the curved space-time, 
the $R^2$ term plays a role of correction of the higher curvature to the Einstein gravity or it contributes to the energy-momentum tensor as matter.

\subsection{$F(R)$ gravity with the quantum anomaly}

The action describing $F(R)$ gravity is given by 
\begin{eqnarray}
S \Eqn{=} \int d^4 x \sqrt{-g}\,
\frac{F(R)}{2\kappa^2} 
+ S^{(\mathrm{QA})}\,, 
\label{eq:3.4} \\ 
F(R) \Eqn{\equiv} 
R+2\kappa^2 \left(\frac{\alpha_4 \bar{N}^2}{192\pi^2}\right) R^2 + f(R) \,, 
\label{eq:3.5}
\end{eqnarray}
where $S^{(\mathrm{QA})}$ is the action of the quantum anomaly and $f(R)$ is a function of $R$. For the original Starobinsky inflation~\cite{Staro}, 
$f(R) = 0$. {}From the action in Eq.~(\ref{eq:3.4}), the gravitational field equation reads  
\begin{eqnarray} 
R_{\mu\nu}-\frac{1}{2}g_{\mu\nu} R \Eqn{=} 
\kappa^2 \left\langle T_{\mu\nu}^{(\mathrm{QA})} \right\rangle 
+ \kappa^2 \left(\frac{\alpha_4 \bar{N}^2}{48\pi^2}\right) 
\left( -R R_{\mu\nu} + \frac{1}{4} R^2 g_{\mu\nu} 
+ \nabla_\mu\nabla_\nu R - g_{\mu\nu}\Box R^2 \right)
\nonumber\\
&&
{}-f_R(R)\left(R_{\mu\nu}-\frac{1}{2}Rg_{\mu\nu}\right)
+\frac{1}{2}g_{\mu\nu}\left( f(R)-R f_R(R) \right) 
\nonumber\\
&&
{}+\left(\nabla_{\mu}\nabla_{\nu}-g_{\mu\nu}\Box \right)f_R(R)\,,
\label{eq:3.6}
\end{eqnarray}
where $f_R (R) \equiv d f(R)/d R$. 
The trace of this equation is expressed as 
$R=-\kappa^2 \left\{  \alpha_1 \mathcal{W}-\alpha_2 \mathcal{G} 
-\left[\alpha_4 \bar{N}^2/\left(16\pi^2\right) \right] 
\Box R \right\} -2f(R)+R f_R(R)+3\Box f_R(R)$. 

In the FLRW background, 
the gravitational field equations are represented as 
\begin{equation}
\frac{3}{\kappa^{2}}H^{2} = \rho_\mathrm{eff}\,, 
\quad 
-\frac{1}{\kappa^{2}} \left( 2\dot H+3H^{2} \right)  
= P_\mathrm{eff}\,. 
\label{eq:3.8}
\end{equation}
Here, $\rho_\mathrm{eff}$ and $P_\mathrm{eff}$ are the effective energy density and pressure and they obey the equation of the conservation low 
$\dot\rho_{\text{eff}}+3H \left(\rho_{\text{eff}}+P_{\text{eff}}\right)=0$. 
Their expressions are given by  
\begin{eqnarray}
\hspace{-15mm}
\rho_\mathrm{eff} \Eqn{\equiv} \rho^{(\mathrm{QA})}
+\frac{1}{2\kappa^2}\left(R f_R(R)-f(R)-6H^2 f_R(R)-6H \dot f_R(R)\right)\,,  
\label{eq:3.9}\\
\hspace{-15mm}
P_\mathrm{eff} \Eqn{\equiv} P^{(\mathrm{QA})} 
+ \frac{1}{2\kappa^{2}} \left[
-R f_R(R)+f(R)+(4\dot H+6H^2)f_R(R)+4H\dot f_R(R)+2\ddot f_R(R)
\right]\,,
\label{eq:3.10}
\end{eqnarray}
with the contributions from the quantum anomaly to the effective energy density and pressure 
\begin{eqnarray}
\hspace{-15mm}
\rho^{(\mathrm{QA})} \Eqn{\equiv} \frac{\bar{\rho}}{a^4}+6 \alpha_2 H^4 
-\left(\frac{\alpha_4 \bar{N}^2}{16\pi^2}\right) 
\left(18H^2\dot H+6\ddot H H-3\dot H^2\right) \,, 
\label{eq:3.11} \\ 
\hspace{-15mm}
P^{(\mathrm{QA})} \Eqn{\equiv} \frac{\bar{\rho}}{3a^4} - \alpha_2 \left(6H^4+8H^2\dot H\right)
+\left(\frac{\alpha_4 \bar{N}^2}{16\pi^2}\right) 
\left(9\dot H^2+12H\ddot H+2\dddot H+18H^2\dot H \right) \,, 
\label{eq:3.12} 
\end{eqnarray}
where $\bar{\rho}$ is an integration constant, and 
in deriving these expressions, we have used the conservation equation for $\rho_\mathrm{eff}$ and $P_\mathrm{eff}$. 
The $\bar{\rho}$ term corresponds to the energy density of radiation 
of the quantum state~\cite{Haw}. 
In what follows, we take $\bar{\rho} = 0$ because at the inflationary stage 
around the Planck scale, the contribution of radiation can be neglected 
in comparison with that of the quantum anomaly as well as that of 
deviation of modified gravity from general relativity.

\subsection{de Sitter solutions by the trace anomaly}

We consider the case that $f(R)$ is given by an exponential form~\cite{Cognola:2008,Linder:2009,Bamba:2010ws}
\begin{equation}
f(R)=-2\Lambda_\mathrm{c} \left[1- \exp\left(-\frac{R}{R_\mathrm{c}}\right)\right]\,, 
\label{eq:3.13}
\end{equation}
where $\Lambda_\mathrm{c} \, (>0)$ and $R_\mathrm{c} \, (>0)$ are positive constants. 
For $R/R_\mathrm{c} \ll 1$, i.e., in the late-time (e.g., present) universe, $f(R)$ approaches zero, and therefore our model becomes equivalent to $R^2$ gravity with the quantum anomaly. 
While for $R/R_\mathrm{c} \gg 1$, namely, in the early universe such as the inflationary stage, the term of $\Lambda_\mathrm{c}$ plays a role of the cosmological constant. When we expand the exponential term as $\exp\left(-R/R_\mathrm{c}\right) = 1 - R/R_\mathrm{c} + O((R/R_\mathrm{c})^2)$, these terms make 
the de Sitter solution realized by the trace anomaly unstable. 
This property can lead to the graceful exit from inflation.  

We derive the de Sitter solution at the inflationary stage. 
In the limit $R/R_\mathrm{c} \gg 1$, from Eq.~(\ref{eq:3.13}) we get 
$f(R)\approx-2\Lambda_\mathrm{c}$. 
For this limit, the Friedmann (first) equation in (\ref{eq:3.8}) becomes 
$\left(3/\kappa^2\right) H^2 \approx 6\alpha_2 H^4 
-3\alpha_4 \bar{N}^2/\left(16\pi^2\right)
\left(6 H^2\dot{H}+2\ddot{H} H -\dot{H}^2\right)
+\left(\Lambda_\mathrm{c}/\kappa^2\right)$. 
Solving this equation, we acquire the de Sitter solution 
\begin{equation}
H_\mathrm{de \,\, Sitter} = \sqrt{ \frac{1}{4\alpha_2 \kappa^2}\left[1\pm\sqrt{1-\frac{8\Lambda_\mathrm{c}\alpha_2\kappa^2}{3}}\right] }\,, 
\label{eq:3.14}
\end{equation}
where we impose the condition $\Lambda_\mathrm{c} < 3/\left(8\alpha_2\kappa^2\right)$ with $\alpha_2 > 0$ so that the solution can be real. 

We examine the instability condition of the de Sitter solution. 
If the de Sitter solution describes inflation, 
it should be unstable because inflation has to end. 
We represent the perturbation as $H=H_\mathrm{de \,\, Sitter} + \delta H(t)$, 
where $\left| \delta H(t) \right| \ll 1$. 
By combining it with the Friedmann equation, we obtain 
$\delta \ddot{H}(t) + 3H_\mathrm{de \,\, Sitter} \delta \dot{H}(t) 
= - \left[\alpha_4 \bar{N}^2/\left(16\pi^2\right)\right] 
\left[\left(1/\kappa^2 \right) -4\alpha_2 H_\mathrm{de \,\, Sitter}^2 
\right] \delta H(t)$, where we have neglected the terms 
proportional to $\exp \left(-R/R_\mathrm{c}\right)$ in Eq.~(\ref{eq:3.13}) 
because the stability of the solution is only related to 
$\Lambda_\mathrm{c}$. The solution for $\delta H(t)$ is written as 
$\delta H(t)= \bar{H} \exp\left(\lambda_\pm t\right)$, 
where $\bar{H}$ is a constant and 
$\lambda_\pm \equiv \left(-3H_\mathrm{de \,\, Sitter} \pm \sqrt{D} \right)/2$ 
(the subscriptions $\pm$ of $\lambda_\pm$ correspond to the signs of 
$\pm$ in the r.h.s.). 
The de Sitter solutions are unstable (and adopted to describe the inflation) 
only if the value of $\lambda_+$ is a real and positive number, namely, 
\begin{equation} 
D \equiv  
9H_\mathrm{de \,\, Sitter}^2
-\frac{64\pi^2}{\alpha_4 \bar{N}^2} J>0\,, 
\quad 
J \equiv  \frac{1}{\kappa^2}-4\alpha_2 H_\mathrm{de \,\, Sitter}^2 >0\,,
\label{3.15}
\end{equation}
where $\alpha_2 >0$ and $\alpha_4 >0$ have been used.

\subsection{Trace-anomaly driven inflation}

We investigate the observable quantities of 
the spectral index $n_\mathrm{s}$ of the power spectrum for the 
the scalar mode of the density perturbations and 
the tensor-to-scalar ratio $r$ 
in the trace-anomaly driven inflation in exponential gravity, 
namely, inflation is described by the de Sitter solutions 
in Eq.~(\ref{eq:3.14}). 
In the slow-roll inflation, 
for the exponential form of $f(R)$ in (\ref{eq:3.13}), 
we have
\begin{equation}
\epsilon \approx 
\frac{u^2}{N_e^2}\frac{\e^{-u} \Lambda_\mathrm{c} \alpha_2 \kappa^2 \left(u+2\right)}{\left(1-8\Lambda_\mathrm{c} \alpha_2 \kappa^2/3 \right)}\ll 1\,,
\quad 
\left|\eta\right| \approx \left|-\frac{u}{2N_e}\right|\ll 1\,,
\label{eq:3.16}
\end{equation}
where $1 \ll u \leq \left(1+\sqrt{1-y}\right)/\left(1-\sqrt{1-y}\right)$ with 
$y \equiv 8\Lambda_\mathrm{c} \alpha_2 \kappa^2/3$. 
As a result, we acquire 
\begin{eqnarray}
\Delta_{\mathcal R}^2 \Eqn{=} \frac{1}{32\pi^2 \alpha_2 \epsilon}\left(1+\sqrt{1-\frac{8\Lambda_\mathrm{c} \alpha_2 \kappa^2}{3}}\right)\,, 
\label{eq:3.17}\\ 
n_\mathrm{s} -1 \Eqn{=} -\frac{u}{N_e}-\frac{6 u^2}{N_e^2}\frac{\e^{-u} \Lambda_\mathrm{c} \alpha_2 \kappa^2 \left(u+2\right)}{\left(1-8\Lambda_\mathrm{c} \alpha_2 \kappa^2/3\right)}\,,
\label{eq:3.18}\\ 
r \Eqn{=} \frac{16 u^2}{N_e^2}\frac{\text{e}^{-u} \Lambda_\mathrm{c} \alpha_2 \kappa^2 \left(u+2\right)}{\left(1-8\Lambda_\mathrm{c} \alpha_2 \kappa^2/3\right)}\,.
\label{eq:3.19}
\end{eqnarray}
For $u=3$, $\Lambda_\mathrm{c} \alpha_2 \kappa^2 = 0.125$, and $N_e =76$, 
we acquire $n_\mathrm{s} = 0.960$ and $r=1.20 \times 10^{-3}$. 
Consequently, the trace-anomaly driven inflation in exponential gravity 
can explain the Planck results.

\section{$R^2$ gravity in loop quantum cosmology (LQC)} 

In this section, 
we review $R^2$ gravity and its cosmological dynamics in LQC with the holonomy corrections along the investigations in Ref.~\cite{Amoros:2014tha}\footnote{For LQC in teleparallelism, 
finite-time future singularities~\cite{Nojiri:2005sx,Bamba:2008ut} 
have been examined in Ref.~\cite{Bamba:2012ka}, and 
bouncing behaviors have also been studied in Refs.~\cite{Amoros:2013nxa,Haro:2014wha}.}.

\subsection{$F(R)$ gravity in LQC}

We explain $F(R)$ gravity in the framework of LQC~\cite{Zhang:2011vi,Zhang:2011qq,Zhang:2012ta}. 
We consider the Einstein frame as in usual LQC only for the FLRW background 
with its spatially flatness~\cite{Gupt:2011jh}. In this case, 
the relation $\{\hat{\beta}_\mathrm{LQC},\hat{V}_\mathrm{volume}\}=\gamma_\mathrm{BI}/2$ is satisfied~\cite{Singh:2009mz}. 
Here, $\{\,\}$ denotes the Poisson bracket in terms of 
classical variables 
$\hat{\beta}_\mathrm{LQC} \equiv \gamma_\mathrm{BI} \hat{H}$ with 
$\gamma_\mathrm{BI}$ the Barbero-Immirzi parameter and the volume 
$\hat{V}_\mathrm{volume} \equiv \hat{a}^3$, where 
$\hat{a} = \sqrt{F_R} a$. 
Moreover, 
$\hat{\beta}_\mathrm{LQC}$ and $\hat{V}_\mathrm{volume}$ 
are canonically conjugated quantities with each other, and 
these variables are the unique combination for a loop quantization~\cite{Corichi:2008zb}. 
We note that the hat shows the quantities in the Einstein frame. 

It is necessary to take the Hilbert space, in which the quantum states 
are described by (almost) periodic functions, so that the property of 
the discrete space can be included. 
For this purpose, we use the Hamiltonian with the general holonomy 
corrections~\cite{abl03,Ashtekar:2006wn}. Namely, we introduce $\lambda_\mathrm{LQC} \equiv 
\sqrt{\left(\sqrt{3}/2\right)\gamma_\mathrm{BI}}$ and 
replace $\hat{\beta}_\mathrm{LQC}$ with $\sin\left(\lambda_\mathrm{LQC} \hat{\beta}_\mathrm{LQC} \right)/\lambda_\mathrm{LQC}$ 
in the Hamiltonian 
$\hat{\mathcal{H}} = 
-3\left(\hat{\beta}_\mathrm{LQC}^2/\gamma_\mathrm{BI}^2\right) \hat{V}_\mathrm{volume}+\hat{V}_\mathrm{volume}\left[\left(1/2\right)(d\hat{\varphi}/d\hat{t})^2+V(\hat{\varphi})\right]=0$ with $d\hat{t} = \sqrt{F_R} dt$ 
by retaining $\{\hat{\beta}_\mathrm{LQC},\hat{V}_\mathrm{volume}\}=\gamma_\mathrm{BI}/2$~\cite{he10,Dzierzak:2009ip,Bojowald:2008ik}. 
As a result, we acquire the novel Hamiltonian 
$\hat{\mathcal{H}}_\mathrm{LQC}$. 
With the Hamiltonian equation 
$d \hat{V}_\mathrm{volume}/d \hat{t} =\{ \hat{V}_\mathrm{volume}, 
\hat{\mathcal{H}}_\mathrm{LQC} \}$ and the Hamiltonian constraint 
$\hat{\mathcal{H}}_\mathrm{LQC} =0$, 
the Friedmann equation with the holonomy corrections reads~\cite{Singh:2009mz}
\begin{equation}
\hat{H}^2=\frac{1}{3}\hat{\rho}\left(1-\frac{\hat{\rho}}{\hat{\rho}_\mathrm{critical}}\right)\,. 
\label{eq:4.1}
\end{equation}
Here, $\hat{\rho}$ is the energy density of matter, and 
$\hat{\rho}_\mathrm{critical} \equiv 3/\left(\lambda_\mathrm{LQC} \gamma_\mathrm{BI} \right)^2$ is the critical energy density.

\subsection{$R^2$ gravity in LQC}

For $R^2$ gravity, whose action is given by 
$S = \int d^4 x \sqrt{-g} \left[F(R)/\left(2\kappa^2\right) \right]$ 
with $F(R) = R + \alpha_\mathrm{S} \kappa^2 R^2$, 
there appears curvature singularities 
in the early universe. 
In what follows, when we consider $R^2$ gravity, we analyze this action. 
On the other hand, for $R^2$ gravity in the context of LQC, it is possible that no singularity happens. We show this point below. It follows from Eq.~(\ref{eq:4.1}) that 
$0\leq \hat {\rho}\leq \hat{\rho}_\mathrm{critical}$ and 
$-\sqrt{\hat{\rho}_\mathrm{critical}/12}\leq \hat{H}\leq \sqrt{\hat{\rho}_\mathrm{critical}/12}$. In addition, for $R^2$ gravity, 
we have $V(\hat{\varphi}) = \left[1/\left(8\alpha_\mathrm{S} \kappa^2\right)\right] \left( 1-\exp\left(-\sqrt{2/3} \kappa \hat{\varphi} \right) \right)^2$ 
in the Einstein frame. 
Therefore, we find 
$0\leq \left(d\hat{\varphi}/d\hat{t}\right)^2 \leq 2\hat{\rho}_\mathrm{critical}$ and 
$0\leq V(\hat{\rho})\leq \hat{\rho}_\mathrm{critical}$. 
For LQC, the Raychaudhuri equation becomes 
$d\hat{H}/d\hat{t} = -\left(1/2\right)\left(d\hat{\varphi}/d\hat{t}\right)^2\left(1-2\hat{\rho}/\hat{\rho}_\mathrm{critical}\right)$. With this equation, we find 
$\left| d\hat{H}/d\hat{t} \right| \leq \left(1/2\right)\left(d\hat{\varphi}/d\hat{t}\right)^2 \leq \hat{\rho}_\mathrm{critical}$, and thus, it follows from this relation that $\left|\hat{R}\right| \leq 7\hat{\rho}_\mathrm{critical}$. 

Furthermore, the potential $V(\hat{\varphi})$ in the Einstein frame described above obeys 
$d V(\hat{\varphi})/d \hat{\varphi} = \left(1/F_R\right)\sqrt{V(\hat{\varphi})/\left(3\alpha_\mathrm{S}\kappa^2 \right)}$. Substituting this equation into 
the relation between $R$ and $\hat{R}$ as 
$R = F_R \left[\hat{R}+ \left(d\hat{\varphi}/d\hat{t}\right)^2 + 
\sqrt{6} \left(d V(\hat{\varphi})/d \hat{\varphi}\right) \right]$ 
and using the relation $1-2\alpha_\mathrm{S}\kappa^2 \left[\hat{R}+\left(d\hat{\varphi}/d\hat{t}\right)^2 \right] \geq 1-18\alpha_\mathrm{S} \kappa^2 \hat{\rho}_\mathrm{critical}$, we acquire
\begin{equation}
\left|R\right|\leq \frac{1}{1-18\alpha_\mathrm{S} \kappa^2 \hat{\rho}_\mathrm{critical}} \left(18\hat{\rho}_\mathrm{critical} + \sqrt{\frac{2\hat{\rho}_\mathrm{critical}}{\alpha_\mathrm{S}\kappa^2}}\right)\,.
\label{eq:4.2}
\end{equation}
For $\alpha_\mathrm{S} \kappa^2<1/\left(18\hat{\rho}_\mathrm{critical}
\right)$, the absolute value of $R$ is bounded as in Eq.~(\ref{eq:4.2}). 
In addition, the relation between $H$ and $\hat{H}$ is represented as 
$H = \sqrt{F_R} \left[ \hat{H} -\left(1/\sqrt{6}\right)\left(d\hat{\varphi}/d\hat{t}\right) \right]$. Accordingly, $\left|H\right|$ is bounded. 
As a result, $\left|\dot{H}\right|=\left(1/6\right)\left|R-12H^2\right|$ 
does not diverge. Thus, there does not appear any singularity in 
$R^2$ gravity for LQC.

\subsection{Loop quantum $R^2$ gravity in the Einstein frame}

We further analyze the dynamics of $R^2$ gravity 
(i.e, $F(R) = R + \alpha_\mathrm{S} \kappa^2 R^2$) 
in the Einstein frame, where equations become simpler than those in 
the Jordan frame. 
The equation of motion for $\hat{\varphi}$ is expressed by 
$d^2\hat{\varphi}/d\hat{t}^2 + 3\hat{H} \left( d\hat{\varphi}/d\hat{t} \right) 
+ dV(\hat{\varphi})/d\hat{\varphi} = 0$ 
with $V(\hat{\varphi}) = \left[1/\left(8\alpha_\mathrm{S} \kappa^2\right)\right] \left( 1-\exp\left(-\sqrt{2/3} \kappa \hat{\varphi} \right) \right)^2$. 
We introduce a variable 
$\hat{\Psi} \equiv \exp\left(\sqrt{2/3} \kappa \hat{\varphi} \right)$. {}From the equation of motion for $\hat{\varphi}$, we obtain 
\begin{equation} 
\frac{d^2\hat{\Psi}}{d \hat{t}^2} \hat{\Psi} - \left(\frac{d\hat{\Psi}}{d \hat{t}}\right)^2 + 3\hat{H} \frac{d\hat{\Psi}}{d \hat{t}} \hat{\Psi} 
+\frac{1}{6\alpha_\mathrm{S} \kappa^2}\left(\hat{\Psi}-1\right)=0\,.
\label{eq:4.3}
\end{equation}
Since this equation is invariant under the transformation 
$(\hat{H}, \hat{t}) \to (-\hat{H}, -\hat{t})$, 
the solution orbit draws the symmetric trajectories 
of the expansion and contraction phases on 
the $(\hat{\Psi}, d\hat{\Psi}/d\hat{t})$ plane 
in terms of the $d\hat{\Psi}/d\hat{t}=0$ axis. 
That is, if there is a trajectory 
$(\hat{\Psi}(\hat{t}), d\hat{\Psi}(\hat{t})/d\hat{t})$ 
for the contraction phase ($\hat{H} <0$), 
we have a trajectory $(\hat{\Psi}(-\hat{t}), -d\hat{\Psi}(-\hat{t})/d\hat{t})$ 
for the expansion phase ($\hat{H} >0$).

In addition, the energy density is expressed as 
\begin{equation} 
\hat{\rho}=\frac{3}{4\hat{\Psi}^2}\left[ 
\left(\frac{d\hat{\Psi}}{d \hat{t}}\right)^2 
+ \frac{1}{6\alpha_\mathrm{S} \kappa^2}\left(\hat{\Psi}-1\right)^2 
\right]\,. 
\label{eq:4.4}
\end{equation}
This implies that $\hat{H} = 0$ at $(\hat{\Psi}(\hat{t}), d\hat{\Psi}(\hat{t})/d\hat{t}) = (1,0)$. For $\hat{\rho} = \hat{\rho}_\mathrm{critical}$, we find 
\begin{equation} 
\frac{\left(d\hat{\Psi}/d \hat{t}\right)^2}{
4\hat{\rho}_\mathrm{critical}/\left(3A\right)} +
\frac{\left(\hat{\Psi}-1/A\right)^2}{8\alpha_\mathrm{S} \kappa^2 
\hat{\rho}_\mathrm{critical}/A^2} = 1\,,
\label{eq:4.5}
\end{equation}
with $A \equiv 1-8\alpha_\mathrm{S} \kappa^2 \hat{\rho}_\mathrm{critical}$. 
This depicts an ellipse for $A>0$, a parabola for $A=0$, and 
a hyperbola for $A<0$. 
There exists only the critical point at $(\hat{\Psi}(\hat{t}), d\hat{\Psi}(\hat{t})/d\hat{t}) = (1,0)$, where $\hat{\rho}=0$. All of the trajectories start from this point and come back to it. Therefore, it corresponds to both the beginning and end points of the universe. 

As a consequence, thanks to the holonomy corrections, in the Einstein frame, 
the bounces can occur when $\hat{\rho}=\hat{\rho}_\mathrm{critical}$. The universe evolves from the contraction phase ($\hat{H} <0$). 
Its trajectory oscillates near the critical point $(\hat{\Psi}(\hat{t}), d\hat{\Psi}(\hat{t})/d\hat{t}) = (1,0)$ and the oscillatory amplitude becomes large. 
Eventually, the trajectory approaches the line $\hat{\rho}=\hat{\rho}_\mathrm{critical}$ and the bounce happens. After that, the expansion phase 
($\hat{H} >0$) begins and the trajectory goes back to the critical point $(\hat{\Psi}(\hat{t}), d\hat{\Psi}(\hat{t})/d\hat{t}) = (1,0)$ with its oscillating behavior.  

Next, we explore the slow-roll inflation. 
With the slow-roll approximations $\left(d\hat{\varphi}/d\hat{t}\right)^2 \ll V(\hat{\varphi})$ and $\left| d^2\hat{\varphi}/d\hat{t}^2 \right| \ll \left| 3\hat{H} \left(d\hat{\varphi}/d\hat{t}\right) \right|$, the Friedmann equation with the holonomy corrections and equation of motion for $\hat{\varphi}$ 
are written as 
$\hat{H}^2 = \left(1/3\right)V(\hat{\varphi})\left(1-V(\hat{\varphi})/\hat{\rho}_\mathrm{critical} \right)$ and 
$3\hat{H} \left(d\hat{\varphi}/d\hat{t}\right) + 
d V(\hat{\varphi}) / d \varphi = 0$, respectively. 
The number of $e$-folds during inflation is derived by 
$\hat{N}_e \equiv \int_{\hat{t}_\mathrm{i}}^{\hat{t}_\mathrm{f}} 
\hat{H} d\hat{t} = 
\int_{\hat{\varphi}_\mathrm{i}}^{\hat{\varphi}_\mathrm{f}}
\hat{H}/\left( d\hat{\varphi}/d\hat{t} \right) d\hat{\varphi} 
\approx 
\int_{\hat{\varphi}_{e}}^{\hat{\varphi}_{i}} 
\left[V(\hat{\varphi})/\left(dV(\hat{\varphi})/d\hat{\varphi}\right)\right] 
\left(1-V(\hat{\varphi})/\hat{\rho}_\mathrm{critical}
\right) d\hat{\varphi}$. For $V(\hat{\varphi}) = \left[1/\left(8\alpha_\mathrm{S} \kappa^2\right)\right] \left( 1-\exp\left(-\sqrt{2/3} \kappa \hat{\varphi} \right) \right)^2$, we find 
$\hat{N}_e \approx \left(3/4\right) 
\exp\left( \sqrt{2/3} \kappa \hat{\varphi}_\mathrm{i} \right)$. 
Moreover, for $\hat{N}_e \gg 1$, 
the slow-roll parameters are described as 
\begin{equation}
\hat{\epsilon} \approx \frac{3}{4\hat{N}_e^2}\frac{\left[1-1/\left(4\alpha_\mathrm{S} \kappa^2 \hat{\rho}_\mathrm{critical} \right)\right]}{\left[1-1/\left( 8\alpha_\mathrm{S} \kappa^2 \hat{\rho}_\mathrm{critical} \right) \right]^2}\,, 
\quad 
\hat{\eta} \approx -\frac{1}{\hat{N}_e} \frac{1}{
\left[1-1/\left( 8\alpha_\mathrm{S} \kappa^2 \hat{\rho}_\mathrm{critical} \right) \right]}\,.
\label{eq:4.6}
\end{equation}
By using these expressions, 
the spectral index of scalar mode of the density perturbations 
$\hat{n}_\mathrm{s} -1 =  -6\hat{\eta}+2\hat{\eta}$ 
and the tensor-to-scalar ratio $\hat{r} = 16 \hat{\epsilon}$ become
\begin{eqnarray}
\hat{n}_\mathrm{s} -1  
\Eqn{\approx} -\frac{2}{\hat{N}_e}\frac{1}{\left[1-1/\left( 8\alpha_\mathrm{S} \kappa^2 \hat{\rho}_\mathrm{critical} \right) \right]}\,, 
\label{eq:4.7}\\ 
\hat{r} \Eqn{\approx} \frac{12}{\hat{N}_e^2} 
\frac{\left[1-1/\left(4\alpha_\mathrm{S} \kappa^2 \hat{\rho}_\mathrm{critical} \right)\right]}{\left[1-1/\left( 8\alpha_\mathrm{S} \kappa^2 \hat{\rho}_\mathrm{critical} \right) \right]^2}\,. 
\label{eq:4.8}
\end{eqnarray}
In the limit $\hat{\rho}_\mathrm{critical} \to \infty$, 
these expressions of $\hat{n}_\mathrm{s}$ and $\hat{r}$ become equivalent to those for pure $R^2$ gravity without the holonomy corrections, i.e., the original Starobinsky inflation model. 
For instance, if $\hat{N}_e = 68.0$ and $8\alpha_\mathrm{S} \kappa^2 \hat{\rho}_\mathrm{critical} = 8.50$, 
we acquire $\hat{n}_\mathrm{s} = 0.967$ and $\hat{r} = 2.55 \times 10^{-3}$. 
Thus, in an $R^2$ gravity model with the holonomy corrections in the context 
of LQC, the spectral index of scalar mode of the density perturbations 
and the tensor-to-scalar ratio can be compatible with 
the Planck data. 
We mention that in the Starobinsky inflation model, 
for $\hat{N}_e = 60.0$ ($68.0$), we have $\hat{n}_\mathrm{s} = 0.967$ ($0. 971$) and $\hat{r} = 3.33 \times 10^{-3}$ ($2.60 \times 10^{-3}$).

\subsection{Loop quantum $R^2$ gravity in the Jordan frame}

Furthermore, in the Jordan frame, 
we analyze the cosmological behaviors in $R^2$ gravity for LQC. 
For the comparison with the consequences in the Einstein frame, 
the point on the $(\hat{\Psi}, d\hat{\Psi}/d\hat{t})$ plane 
at which the bounce occurs, i.e., $H$ becomes zero. {}From 
$H = \sqrt{F_R} \left[ \hat{H} -\left(1/\sqrt{6}\right)\left(d\hat{\varphi}/d\hat{t}\right) \right]$, if $H=0$, we get $\hat{H}^2 = \left(1/4\right) \left(d\hat{\Psi}/d\hat{t}\right)^2/\hat{\Psi}^2$. This equation leads to 
\begin{equation}
\frac{\left(d\hat{\Psi}/d \hat{t}\right)^2}{
\hat{\rho}_\mathrm{critical}/\left(12B_\pm\right)} +
\frac{\left(\hat{\Psi}-C_\pm/B_\pm \right)^2}{2\alpha_\mathrm{S} \kappa^2 
\hat{\rho}_\mathrm{critical}/B_\pm^2} = 1\,,
\label{eq:4.9} 
\end{equation}
with $B_\pm \equiv 1 \pm \sqrt{8\alpha_\mathrm{S} \kappa^2 
\hat{\rho}_\mathrm{critical}}$ and $C_\pm \equiv 1 \pm \sqrt{2\alpha_\mathrm{S} \kappa^2 \hat{\rho}_\mathrm{critical}}$, where
the subscription $\pm$ in $B_\pm$ and $C_\pm$ corresponds to the sign $\pm$. 
The case of $+$ sign is for $0< \hat{\Psi} < 1$. 
In this case, this curve draws an ellipse for $B_+>0$, a parabola for $B_+=0$, and an hyperbola for $B_+<0$. On the other hand, the case of $-$ sign is for $\hat{\Psi} > 1$. In this case, it shows an ellipse. 
If the trajectory intersects this curve in the Einstein frame, 
the bounce happens in the Jordan frame. 
At the bouncing point, the relation 
$\hat{H} = \left(1/2\right) \left(d\hat{\Psi}/d\hat{t}\right)/\hat{\Psi}$ 
has to be met. 

In the Einstein frame, 
the universe first contracts and finally expands at the critical point 
$(\hat{\Psi}(\hat{t}), d\hat{\Psi}(\hat{t})/d\hat{t}) = (1,0)$. 
With the relation between $H$ and $\hat{H}$ and its time derivative 
\begin{eqnarray}
H \Eqn{=} \sqrt{\hat{\Psi}}\left( \hat{H}
-\frac{1}{2} \frac{d\hat{\Psi}/d\hat{t}}{\hat{\Psi}} \right)\,,
\label{eq:4.10}\\
\dot{H} \Eqn{=} \frac{1}{2} \frac{d\hat{\Psi}}{d\hat{t}}  
\left( \hat{H}
-\frac{1}{2} \frac{d\hat{\Psi}/d\hat{t}}{\hat{\Psi}} \right) 
+ \hat{\Psi} \left[
\frac{d\hat{H}}{d\hat{t}} -\frac{1}{2}
\frac{d}{d\hat{t}} \left( 
\frac{d\hat{\Psi}/d\hat{t}}{\hat{\Psi}} \right)
\right]\,,
\label{eq:4.11} 
\end{eqnarray}
it is seen that in the Jordan frame, 
the universe begins and ends at the point $(H, \dot{H}) = (0,0)$. 
It should be emphasized that the holonomy corrections yield the bounce 
in the Jordan frame, and hence, if they are absent, a singularity 
appears at the early stage of the universe.

\section{Bouncing cosmology in $F(R)$ gravity}

In this section, we review the cosmological bounce from $F(R)$ gravity. 
Especially, we present the consequences found in Refs.~\cite{Odintsov:2014gea,Odintsov:2015uca}. The bouncing behaviors in various modified gravity theories have also been investigated in Refs.~\cite{deHaro:2014tla,Elizalde:2014uba,deHaro:2014kxa,Bamba:2014zoa,Bamba:2014mya,Bamba:2013fha}. 
We show that it is possible to reconstruct an $F(R)$ gravity theory in which 
the matter bounce can happen in the framework of LQC. 

For the Friedmann equation with the holonomy corrections (\ref{eq:4.1}), 
the energy density of matter can be represented as
$\rho = \bar{\rho}_\mathrm{m}/\left[\left(3/4\right) t^2+1\right]$ 
with $\bar{\rho}_\mathrm{m}$ a constant. 
In this case, the scale factor and the Hubble parameter read~\cite{Haro:2014wha,deHaro:2014kxa} 
\begin{equation}
a(t)=\left (\frac{3}{4} \bar{\rho}_\mathrm{m} t^2+1\right )^{1/3}\,,
\quad 
H(t)=\frac{\left(1/2\right) \bar{\rho}_\mathrm{m} t}{\left(3/4\right) \bar{\rho}_\mathrm{m} t^2+1} \,.
\label{eq:5.1} 
\end{equation}
For these expressions, 
the solution of Eq.~(\ref{eq:2.10}) yields the form of $F(R)$ to realize 
the matter bounce described by $a$ and $H$ in (\ref{eq:5.1}) for LQC. 
We solve Eq.~(\ref{eq:5.1}), and consequently acquire 
\begin{equation}
F(R)= \mathcal{I}_1 R+\mathcal{I}_2R^{1/2}\,, 
\label{eq:5.2} 
\end{equation}
where $\mathcal{I}_1$ and $\mathcal{I}_2$ are constants. 
We can set $\mathcal{I}_1 = 1$, so that the Einstein-Hilbert term 
should be included. 

In Ref.~\cite{Odintsov:2015uca}, the reconstruction of various modified gravity theories including $F(R)$, $F(\mathcal{G})$, and $F(T)$ gravity theories has been performed, where $F(\mathcal{G})$ is an arbitrary function of the Gauss-Bonnet invariant $\mathcal{G}$, to describe the two-times bouncing phenomena, called super-bounce~\cite{Koehn:2013upa,Cai:2012va}, and the ekpyrotic scenario~\cite{Khoury:2001wf} in the context of LQC.

\section{Conclusions}

In the present paper, we have reviewed inflationary models in modified gravity theories such as $F(R)$ gravity including $R^2$ gravity with extended terms 
so that we can generalize the Starobinsky inflation in $R^2$ gravity 
and derive its important properties to be useful clues to obtain the information on physics in the early universe. 

First, we have studied inflationary cosmology by 
modification terms of gravity, especially, inflation in $F(R)$ gravity, 
by following Ref.~\cite{Sebastiani:2013eqa}. 
The Starobinsky inflation in $R^2$ gravity is considered to be 
the seminal and significant idea of inflationary models in modified gravity theories. We have made the coformal transformation from the Jordan frame (namely, $F(R)$ gravity) to the Einstein frame (i.e., general gravity plus the scalar field theory), and given slow-roll dynamics of inflation in the Einstein 
frame. In addition, we have reconstructed $F(R)$ gravity models, which are 
an extended version of the Starobinsky inflation model in $R^2$ gravity and general relativity with power-law correction terms. 

Second, we have explored the trace-anomaly driven inflation 
in $F(R)$ gravity along the discussions in Ref.~\cite{Bamba:2014jia}. We have first explained the quantum anomaly appearing through the process of the renormalization in four-dimensional space-time. 
We have further discussed $F(R)$ gravity with the quantum anomaly 
and the de Sitter solutions for inflation due to the trace anomaly. 

Third, based on Ref.~\cite{Amoros:2014tha}, we have examined inflation in $R^2$ gravity and the cosmological evolutions for LQC with the holonomy corrections. We have analyzed $R^2$ gravity 
for LQC in both the Einstein and Jordan frames. We have found that in the Jordan frame, owing to the holonomy corrections, the bounce can happen, and accordingly the cosmic singularities can be removed, although such singularities 
appear in ordinary $R^2$ gravity. 

In these three inflationary models, 
we have shown that the spectral index of scalar modes of the density perturbations and the tensor-to-scalar ratio can be 
compatible with the Planck analysis. 

Furthermore, we have presented the recent developments of the bounce cosmology in $F(R)$ gravity obtained in Refs.~\cite{Odintsov:2014gea,Odintsov:2015uca}. It has been performed that an $F(R)$ gravity theory can be reconstructed, where the matter bounce occurs in the context of LQC, thanks to the reconstruction method of $F(R)$ gravity. Recently, the reconstruction of $F(R)$, $F(\mathcal{G})$, and $F(T)$ gravity theories have also been executed, in which the super-bounce (i.e., two-times bounce behaviors) and the ekpyrotic scenario for LQC can be realized. 

In this work, we have concentrated on the accelerating universe from $F(R)$ gravity. Note, however, that it is possible to extend this study for more complicated versions of effective gravity, which comes from quantum gravity. Particularly, it has recently been demonstrated that successful inflation consistent with the Planck data may emerge from multiplicatively-renormalizable higher derivative quantum gravity in Ref.~\cite{Myrzakulov:2014hca}.

\section*{Acknowledgments}

This work was partially supported by the JSPS Grant-in-Aid for Young Scientists (B) \# 25800136 (K.B.) and MINECO (Spain) project FIS2010-15640 and FIS2013-44881 (S.D.O.).



\end{document}